\documentclass[journal]{IEEEtran}

\usepackage{subfigure}
\usepackage{amssymb}
\usepackage{dsfont} 
\usepackage{amsbsy} 
\usepackage{amsmath} 
\usepackage{amsmath, amsthm}
\usepackage[table]{xcolor}
\usepackage{booktabs}
\usepackage{multirow}
\usepackage{eufrak}
\usepackage{mathrsfs}
\usepackage{bbm}
\usepackage{enumerate}

\DeclareMathAlphabet{\mathpzc}{OT1}{pzc}{m}{it}
\theoremstyle{remark}
\newtheorem*{remark}{Remark}
\newtheorem*{assumption}{Assumptions}

\ifCLASSINFOpdf
   \usepackage[pdftex]{graphicx}

\else
  
\fi

\newcommand{\E}[1]{\mathbb{E}\left[ #1 \right]}

\newcommand{\vect}[1]{\mathrm{vec}\left\{ #1 \right\}}

\begin{document}

\title{Partial Diffusion Recursive Least-Squares for Distributed Estimation under Noisy Links Condition}

\author{Vahid Vahidpour,
Amir Rastegarnia,
Azam Khalili, and
Saeid Sanei,~\IEEEmembership{Senior~Member,~IEEE}

%%%%\thanks{Manuscript received XXXX, XXX; revised XXXX, 2016. This work
%%%was supported in part by YYYY. The associate editor coordinating the review of this manuscript and approving it for publication was XXXX XXXX.
%%%
%%%F. A. Author is with the National Institute of Standards and Technology, Boulder, CO 80305 USA (corresponding author to provide phone: 303-555-5555; fax: 303-555-5555; e-mail: author@ boulder.nist.gov). 
%%%S. B. Author, Jr., was with Rice University, Houston, TX 77005 USA. He is now with the Department of Physics, Colorado State University, Fort flins, CO 80523 USA (e-mail: author@lamar.colostate.edu).}
%%%\thanks{Digital Object Identifier XXXXX/XXXXXXX}
\thanks{
V. Vahidpour, A. Rastegarnia, and A. Khalili and  are with the Department of Electrical Engineering, Malayer University, Malayer 65719-95863, Iran (email: v.vahidpour@gmail.com; rastegar@tabrizu.ac.ir; a-khalili@tabrizu.ac.ir).

S. Sanei is with the Department of Computer Science, University of Surrey, Surrey GU2 7XH, UK (email: s.sanei@surrey.ac.uk).}
}

\markboth{DRAFT}%
{Shell \MakeLowercase{\textit{et al.}}: Bare Demo of IEEEtran.cls for Journals}
%\markboth{IEEE TRANSACTIONS ON BIOMEDICAL ENGINEERING,~VOL. X, 2013}%
%{Shell \MakeLowercase{\textit{et al.}}: Bare Demo of IEEEtran.cls for Journals}
%\IEEEpubid{0000--0000/00\$00.00~\copyright~2013 IEEE}

\maketitle

\begin{abstract}
Partial diffusion-based recursive least squares (PDRLS) is an effective method for reducing computational load and power consumption in adaptive network implementation. In this method, each node shares a part of its intermediate estimate vector with its neighbors at each iteration. PDRLS algorithm reduces the internode communications relative to the full-diffusion RLS algorithm. This selection of estimate entries becomes more appealing when the information fuse over noisy links. In this paper, we study the steady-state performance of PDRLS algorithm in presence of noisy links and investigate its convergence in both mean and mean-square senses. We also derive a theoretical expression for its steady-state mean-square deviation (MSD). The simulation results illustrate that the  stability conditions for PDRLS under noisy links are not sufficient to guarantee its convergence. Strictly speaking, considering non-ideal links condition adds a new complexity to the estimation problem for which the PDRLS algorithm becomes unstable and do not converge for any value of the forgetting factor. 
\end{abstract}

\begin{IEEEkeywords}
Adaptive networks, diffusion adaptation, distributed estimation, energy conservation, recursive least-square, partial diffusion, noisy links.
\end{IEEEkeywords}

\IEEEpeerreviewmaketitle

\section{Introduction}
\IEEEPARstart{W}{e} study the problem of distributed estimation over adaptive networks, in which a set of agents are interacted with each other to solve distributed estimation and inference problems in a collaborative manner. There exist several useful techniques for solving such optimization problems in distributed manner that enable adaptation and learning in real-time. They include incremental \cite{Lopes2007,Rabbat2005,Nedic2001}, consensus \cite{Bertrand2011,Stankovic2011,Mateos2009a}, and diffusion \cite{Chen2012,Zhao2012a,Bertrand2011a,Chouvardas2011a,Takahashi2010a,Cattivelli2010a,Lopes2008,Tu2012,Cattivelli2008} strategies. The diffusion strategies are effective methods for performing distributed estimation over adaptive networks.
In the original diffusion strategy, all agents in the network generate their individual intermediate estimates using the data accessible to them locally. Then, the nodes exchange their intermediate estimates to all their immediate neighbors. However, the most expensive part of realizing a cooperative task over a wireless ad hoc network is usually the data communications through radio links. Therefore, it is of practical importance to reduce the amount of internode communications in diffusion strategies while maintaining the benefits of cooperation.

To aim this, various techniques have been proposed, such as choosing a subset of the nodes \cite{Rortveit2010,Lopes2008a,Takahashi2010,Zhao2012}, selecting a subset of the entries of the estimates \cite{Arablouei2014a,Arablouei2014}, and reducing the dimension of the estimates \cite{Sayin2013,Sayin2014,Chouvardas2013}. Among these methods, we focus on the second method in which a subset of the entries are selected in communications.

In all mentioned works, the data are assumed to be exchanged among neighboring nodes without distortion. However, due to the link noise, this assumption may not be true in practice. Some useful results out of studying the effect of link noise during the exchange of weight estimates, already appear for traditional diffusion algorithm \cite{Abdolee2011,Khalili2012a,Khalili2012,Tu2011a}, for incremental case \cite{Khalili2011b,Khalili2011,Khalili2011a}, and for consensus-based algorithms \cite{Kar2009,Mateos2009}.

In partial-diffusion strategies proposed in \cite{Arablouei2014a,Arablouei2014}, the links between nodes are assumed to be ideal. However, the performance of these strategies, due to replacing the unavailable entries by their corresponding ones in each node's own intermediate estimate vector, are strongly affected under noisy information exchange.

The main contributions of this paper can be summarized as follows:
\begin{enumerate}[(i)]
	\item Focusing on \cite{Arablouei2014}, we consider the fact that the weight estimates exchanged among the nodes can be subject to quantization errors and additive noise over communication links. We also consider two different schemes \cite{Arablouei2014a} for selecting the weight vector entries for transmission at each iteration. We allow for noisy exchange just during the two combination steps. It should be noted that since our objective is to minimize the internode communication, the nodes only exchange their intermediate estimates with their neighbors;
	\item Using the energy conservation argument \cite{sayed2011adaptive} we analyze the stability of algorithms in mean and mean square sense under certain statistical conditions;
	\item Stability conditions for PDRLS are derived under noisy links, and it is demonstrated that the steady-state performance of the new algorithm is not as good as that of the PDRLS algorithm with ideal links;
	\item We derive a variance relation which contains an additional terms in comparison to original PDRLS algorithm that represent the effects of noisy links. We evaluate these moments and derived closed-form expressions for mean-square derivation (MSD) to explain the steady-state performance.
	\item It is demonstrated through different examples that the stability conditions for PDRLS under noisy links are not sufficient to guarantee its convergence. Considering noisy links adds a new complexity to estimation problem for which the PDRLS algorithm suffers a significant degradation in steady-state performance for any value of forgetting factor.
	\end{enumerate}
	
The remainder of this paper is organized as follows: In Section II, we formulate the PDLMS under imperfect information exchange. The performance analyses are examined in Section III, where the methods of entry selection matrix are also examined. We provide simulation results in Section IV and draw the conclusions in Section V.

\subsection{Notation} We use the lowercase boldface letters to denote vectors, uppercase boldface letter for matrices, and lowercase plain letters for scalar variables. We also use $(.)^*$ to denote conjugate transposition, $\mathrm{Tr}(.)$ for the trace of its matrix argument, $\otimes$ for the Kronecker product, and $\mathrm{vec}{.}$ for a vector formed by stacking the columns of its matrix argument. We further use $\mathrm{diag}\{\cdots\}$ to denote a (block) diagonal matrix formed from its argument, and $\mathrm{col}\{\cdots\}$ to denote a column vector formed by stacking its arguments on top of each other. All vectors in our treatment are column vectors, with the exception of regression vectors, $\boldsymbol{u}_{k,i}$.

\section{Algorithm Description}
\subsection{Diffusion Recursive Least-Squares Algorithm with Noisy Links}
To begin with, consider a set of $N$ nodes, spatially distributed over some region, aim to identify an unknown parameter vector, $w^{o}\in C^{M\times1}$, in a collective manner. At every time instant $i$, node $k$ collects a measurement $d_k (i)$ that is assumed to be related to an unknown vector by
\begin{equation}
d_k (i)=\boldsymbol{u}_{k,i} w^o+v_k (i)
\label{eq1}
\end{equation}
where $\boldsymbol{u}_k (i)$ is a row vector of length $M$ (the regressor of node $k$ at time $i$) and $v_k (i)$ represent additive noise with zero mean and variance $\sigma^{2}_{\upsilon,k}$ and the unknown vector $w^o$ denotes the parameter of interest.

Collecting the measurements and noise samples of node $k$ up to time $i$ into vector quantities as follows:
\begin{equation}
\boldsymbol{y}_{k,i}=\mathrm{col}\left\{d_k (i),\ldots,d_k (0)\right\}
\label{eq2}
\end{equation}
\begin{equation}
\boldsymbol{H}_{k,i}=\mathrm{col}\left\{\boldsymbol{u}_k (i),\ldots,\boldsymbol{u}_k (0)\right\}
\label{eq3}
\end{equation}
\begin{equation}
\boldsymbol{v}_{k,i}=\mathrm{col}\left\{v_k (i),\ldots,v_k (0)\right\}
\label{eq4}
\end{equation}
the objective is to estimate $w^o$ by solving the following weighted least-squares problem
\begin{equation}
\min_{\boldsymbol{\psi}} \left\|\boldsymbol{y}_{k,i}-\boldsymbol{H}_{k,i}\boldsymbol{\psi}\right\|^{2}_{\Lambda_{i}}
\label{eq5}
\end{equation}
The solution $\boldsymbol{\psi}_{k,i}$ is given by \cite{Sayed2003}
\begin{equation}
\boldsymbol{\psi}_{k,i}=\left(\boldsymbol{H}^{*}_{k,i}\boldsymbol{\Lambda}_{i}\boldsymbol{H}_{k,i}\right)^{-1}\left(\boldsymbol{H}^{*}_{k,i}\boldsymbol{\Lambda}_{i}\boldsymbol{y}_{k,i}\right)
	\label{eq6}
\end{equation}
where $\boldsymbol{\Lambda}_{i}\geq0$ denotes a Hermitian weighting matrix. Common choice for $\boldsymbol{\lambda}_{i}$ is
\begin{equation}
	\boldsymbol{\Lambda}_{i}=\mathrm{diag}\left\{1,\lambda,\ldots,\lambda^{i}\right\}
	\label{eq7}
\end{equation}
where $0\ll\lambda\leq1$ is a forgetting factor whose value is generally very close to one.

It can be verified from the properties of recursive least-squares solutions \cite{Sayed2003,sayed2011adaptive} that
\begin{equation}
	\boldsymbol{H}^{*}_{k,i}\boldsymbol{\Lambda}_{i}\boldsymbol{H}_{k,i}=\lambda\boldsymbol{H}^{*}_{k,i-1}\boldsymbol{\Lambda}_{i-1}\boldsymbol{H}_{k,i-1}+\boldsymbol{u}^{*}_{k,i}\boldsymbol{u}_{k,i}
	\label{eq8}
\end{equation}
\begin{equation}
	\boldsymbol{H}^{*}_{k,i}\boldsymbol{\Lambda}_{i}\boldsymbol{y}_{k,i}=\lambda\boldsymbol{H}^{*}_{k,i-1}\boldsymbol{\Lambda}_{i-1}\boldsymbol{y}_{k,i-1}+\boldsymbol{u}^{*}_{k,i}d_{k}(i)
	\label{eq9}
\end{equation}

Let $\boldsymbol{P}_{k,i}=\left(\boldsymbol{H}^{*}_{k,i}\boldsymbol{\Lambda}_{i}\boldsymbol{H}_{k,i}\right)^{-1}$. Then applying the so-called matrix inversion formula \cite{meyer2000matrix}[36] to \eqref{eq8} the  following recursions for calculating $\boldsymbol{\psi}_{k,i}$ are obtained
\begin{equation}
	\boldsymbol{P}_{k,i}=\lambda^{-1}\left(\boldsymbol{P}_{k,i-1}-\frac{\lambda^{-1}\boldsymbol{P}_{k,i-1}\boldsymbol{u}^{*}_{k,i}\boldsymbol{u}_{k,i}\boldsymbol{P}_{k,i-1}}{1+\lambda^{-1}\boldsymbol{u}_{k,i}\boldsymbol{P}_{k,i-1}\boldsymbol{u}^{*}_{k,i}}\right)
	\label{eq10}
\end{equation}
\begin{equation}
	\boldsymbol{\psi}_{k,i}=\boldsymbol{\psi}_{k,i-1}+\boldsymbol{P}_{k,i}\boldsymbol{u}^{*}_{k,i}\left(d_{k}(i)-\boldsymbol{u}_{k,i}\boldsymbol{\psi}_{k,i-1}\right)
	\label{eq11}
\end{equation}
Since $\boldsymbol{w}_{k,i}$ is a better estimate compared with $\boldsymbol{\psi}_{k,i}$, it is beneficial to replace $\boldsymbol{\psi}_{k,i-1}$ with $\boldsymbol{w}_{k,i-1}$ in \eqref{eq11}
\begin{equation}
	\boldsymbol{\psi}_{k,i}=\boldsymbol{w}_{k,i-1}+\boldsymbol{P}_{k,i}\boldsymbol{u}^{*}_{k,i}\left(d_{k}(i)-\boldsymbol{u}_{k,i}\boldsymbol{w}_{k,i-1}\right)
	\label{eq12}
\end{equation}
Hence, the local estimates are diffused outside of each node's own neighborhood. Then, for every time instant $i$, each node $k$ performs an adaptation step followed by a combination step as follows:
\begin{enumerate}
	\item \textit{Adaptation:} Each node computes an intermediate estimate of $w^{o}$ by \eqref{eq10} and \eqref{eq12}. The resulting pre-estimates are named $\boldsymbol{\psi}_{k,i}$ as in \eqref{eq13}.
	\item \textit{Combination:} The nodes exchange their local pre-estimates with their neighbors and perform a weighted average as in \eqref{eq14} to obtain the estimate $\boldsymbol{w}_{k,i}$ (via so-called spatial update).
\end{enumerate}
\begin{equation}
	\boldsymbol{w}_{k,i}=\sum^{N}_{l=1}a_{lk}\boldsymbol{\psi}_{l,i}
	\label{eq13}
\end{equation}
The scalar $a_{lk}$ is non-negative real coefficient corresponding to the $(l,k)$entries of $N\times N$ combination matrix $A=\{a_{lk}\}$. These coefficients are zero whenever node $l\notin{\mathcal{N}_k}$, where $\mathcal{N}_k$ denotes the neighborhood of node $k$. This matrix is assumed to satisfy the condition:
\begin{equation}
	A^{T}\mathbbm{1}_{N}=\mathbbm{1}_{N}
	\label{eq14}
\end{equation}
where the notation $\mathbbm{1}$ denotes an $N \times 1$ column vector with all its entries equal to one.

The above algorithm only uses the local input-output data, observed by each node, in the adaptation phase. However, in \cite{Cattivelli2008}, a more general algorithm has been proposed in which each node shares its input-output data as well as its intermediate estimate with its neighbors and updates its intermediate estimate using all available data. This is carried out via a convex combination of the update terms induced by each input-output data pair. For the obvious reason of minimizing the communications, here, we will only consider the above-mentioned diffusion RLS algorithm.

We model the noisy data received by node $k$ from its neighbor $l$ as follows:
\begin{equation}
	\boldsymbol{\psi}_{lk,i}=\boldsymbol{\psi}_{l,i}+\boldsymbol{v}^{\left(\psi\right)}_{lk,i}
	\label{eq15}
\end{equation}
where $\boldsymbol{v}^{\left(\psi\right)}_{lk,i}$ $(M\times 1)$ denotes vector noise signal. It is temporally white and spatially independent random process with zero mean and covariance given by $\boldsymbol{R}^{\left(\psi\right)}_{v,lk}$. The quantity $\boldsymbol{R}^{\left(\psi\right)}_{v,lk}$ are all zero if $l\notin{\mathcal{N}_k}$ or when $l=k$. it should be noted that the subscript $lk$ indicates that $l$ is the source and $k$ is the sink, i.e. the flow of information is from $l$ to $k$.
 
Using the perturbed estimate \eqref{eq15}, the combination step in adaptive strategy becomes
\begin{equation}
	\boldsymbol{w}_{k,i}=\sum_{l\in{\mathcal{N}_k}}a_{lk}\boldsymbol{\psi}_{lk,i}
	\label{eq16}
\end{equation}
\subsection{Partial-Diffusion RLS Algorithm under Noisy Information Exchange}
In order to reduce the amount of communication required among the nodes we utilize partial-diffusion strategy proposed in \cite{Arablouei2014a}, to transmit $L$ out of $M$ entries of the intermediate estimates at each time instant where the integer $L$ is fixed and pre-specified. The selection of the to-be-transmitted entries at node $k$ and time instant $i$ can be characterized by an $M\times M$ diagonal entry-selection matrix, denoted by $\boldsymbol{\mathcal{K}}_{k,i}$, that has $L$ ones and $M-L$ zeros on its diagonal. The proposition of ones specifies the selected entries. Multiplication of an intermediate estimate vector by this matrix replaces its non-selected entries with zero.

Rewriting \eqref{eq16} as
%\begin{IEEEeqnarray}{rCl}
%\boldsymbol{w}_{k,i}=a_{kk}\boldsymbol{\psi}_{k,i}+\sum_{l\in{\mathcal{N}_k}\backslash\left\{k\right\}}a_{lk}
%[\boldsymbol{\mathcal{K}}_{l,i}\boldsymbol{\psi}_{lk,i}\nonumber\\
%+\left(\boldsymbol{I}_M-\boldsymbol{\mathcal{K}}_{l,i}\right)\boldsymbol{\psi}_{lk,i}]
%\label{eq14}
%\end{IEEEeqnarray}
\begin{align}
	\boldsymbol{w}_{k,i}&=a_{kk}\boldsymbol{\psi}_{k,i}+\sum_{l\in{\mathcal{N}_k}\backslash\left\{k\right\}}a_{lk}
[\boldsymbol{\mathcal{K}}_{l,i}\boldsymbol{\psi}_{lk,i} \nonumber \\
& \hspace{3cm} +\left(\boldsymbol{I}_M-\boldsymbol{\mathcal{K}}_{l,i}\right)\boldsymbol{\psi}_{lk,i}]
\label{eq17}
	%\boldsymbol{w}_{k,i}&=\nonumber\\
%&a_{kk}\boldsymbol{\psi}_{k,i}+\sum_{l\in{\mathcal{N}_k}\backslash\left\{k\right\}}a_{lk}\left[\boldsymbol{\mathcal{K}}_{l,i}\boldsymbol{\psi}_{lk,i}+\left(\boldsymbol{I}_M-\boldsymbol{\mathcal{K}}_{l,i}\right)\boldsymbol{\psi}_{lk,i}\right]
	%\label{eq17}
\end{align}
Each node requires the knowledge of all entries of its neighbors' intermediate estimate vectors for combination. However, when the intermediate estimates are partially transmitted $(0 < L < M)$, the nodes have no access to the non-communicated entries. To resolve this ambiguity, we let the nodes use the entries of their own intermediate estimates in lieu of ones from the neighbors that have not been communicated, i.e., at node $k$, substitute
\begin{eqnarray}
\left(\boldsymbol{I}_M-\boldsymbol{\mathcal{K}}_{l,i}\right)\boldsymbol{\psi}_{k,i}\ ,\forall{l\in{\mathcal{N}_k}\backslash\left\{k\right\}}
\label{eq18}
\end{eqnarray}
for
\begin{eqnarray}
\left(\boldsymbol{I}_M-\boldsymbol{\mathcal{K}}_{l,i}\right)\boldsymbol{\psi}_{lk,i}\ ,\forall{l\in{\mathcal{N}_k}\backslash\left\{k\right\}}
\label{eq19}
\end{eqnarray}

Based on this approach, we formulate a partial-diffusion recursive least-squares (PDRLS) algorithm under imperfect information exchange using \eqref{eq10} and \eqref{eq12} for adaptation and the following equation for combination:
\begin{eqnarray}
\boldsymbol{w}_{k,i}=\nonumber\\
&&a_{kk}\boldsymbol{\psi}_{k,i}+\sum_{l\in{\mathcal{N}_k}\backslash\left\{k\right\}}a_{lk}[\boldsymbol{\mathcal{K}}_{l,i}\boldsymbol{\psi}_{lk,i}+\left(\boldsymbol{I}_M-\boldsymbol{\mathcal{K}}_{l,i}\right)\boldsymbol{\psi}_{k,i}]\nonumber\\
&&+\boldsymbol{v}^{\left(\psi\right)}_{k,i}
\label{eq20}
\end{eqnarray}
where $\boldsymbol{v}^{(\psi)}_{k,i}$ denotes the aggregate $M\times1$ zero mean noise signal and is introduced as follows:
\begin{equation}
	\boldsymbol{v}^{(\psi)}_{k,i}=\sum_{l\in{\mathcal{N}_k}\backslash\left\{k\right\}}a_{lk}\boldsymbol{\mathcal{K}}_{l,i}\boldsymbol{v}^{\left(\psi\right)}_{lk,i}
	\label{eq21}
\end{equation}
This noise represents the aggregate effect on node $k$ of all selected exchange noises from the neighbors of node $k$ while exchanging the estimates $\left\{\boldsymbol{\psi}_{l,i}\right\}$ during the combination step.

\subsection{Entry Selection Methods}
In order to select $L$ out of $M$ entries of the intermediate estimates of each node at each iteration, the processes we utilized are analogous to the selection processes in stochastic and sequential \textit{partial-update} schemes \cite{Dogancay2008,Godavarti2005,Douglas1997,Treichler1987}. In other words, we use the same schemes as introduced in \cite{Arablouei2014a}. Here, we just review these methods named \textit{sequential} partial-diffusion and \textit{stochastic} partial-diffusion.

In sequential partial-diffusion the entry selection matrices, $\boldsymbol{\mathcal{K}}_{k,i}$, are diagonal matrices:
\begin{equation}
%\left.\begin{aligned}
\boldsymbol{\mathcal{K}}_{k,i}=
\begin{bmatrix}
\kappa_{1,i} & \cdots & 0 \\
\vdots & \ddots & \vdots \\
0 & \cdots & \kappa_{M,i}
\end{bmatrix}
%\end{aligned}
\; ,\kappa_{1,i}=
\begin{cases}
1   & \text{if}\ \ell\in\mathcal{J}_{(i\,\text{{mod}}\,\bar{B})+1}\\
0   & \text{otherwise} 
\end{cases}
\label{eq22}
\end{equation}
with $\bar{B}=\left\lceil M/L\right\rceil$. The number of selection entries at each iteration is limited by $L$. The coefficient subsets $\mathcal{J}_i$ are not unique as long as they obey the following requirements \cite{Dogancay2008}:
\begin{enumerate}
	\item Cardinality of $\mathcal{J}_i$ is between $1$ and $L$;
	\item $\bigcup^{\bar{B}}_{\kappa=1}=\mathcal{S}$\ \text{where}\ $\mathcal{S}=\left\{1,2,\ldots,M\right\}$;
	\item $\mathcal{J}_\kappa \cap \mathcal{J}_\eta=\emptyset$, $\forall\kappa ,\eta\in\left\{1,\ldots,\bar{B}\right\}$ and $\kappa\neq\eta$.
\end{enumerate}

The description of the entry selection matrices, $\boldsymbol{\mathcal{K}}_{k,i}$, in stochastic partial-diffusion is similar to that of sequential one. The only difference is as follows. At a given iteration, $i$, of the sequential case one of the sets $\mathcal{J}_\kappa$, $\kappa=\left\{1,\ldots,\bar{B}\right\}$ is chosen in advance, whereas for stochastic case, one of the sets $\mathcal{J}_\kappa$ is sampled at random from $\left\{\mathcal{J}_{1},\mathcal{J}_{2},\ldots,\mathcal{J}_{\bar{B}}\right\}$. One might ask why these methods are considered to organize these selection matrices. To answer this question, it is worth mentioning that the nodes need to know which entries of their neighbors' intermediate estimates have been transmitted at each iteration. These schemes bypass the need for addressing.
\begin{remark}The probability of transmission for all the entries at each node is equal and expressed as
\begin{equation}
\rho=L/M
\label{eq23}
\end{equation}
\label{re}
\end{remark}
Moreover, the entry selection matrices, $\boldsymbol{\mathcal{K}}_{k,i}$, do not depend on any data/parameter other than $L$ and $M$.
\section{Performance Analysis}
In this section, we analyze the performance of the algorithm \eqref{eq10}, \eqref{eq12} and \eqref{eq20} and show that it is asymptotically unbiased in the mean and converges in the mean-square error sense under some simplifying assumptions. We also provide an expression for mean-square deviation (MSD). We presume that the input data is stochastic in nature and utilize the energy conservation arguments previously applied to LMS-type adaptive distributed algorithms in, e.g., \cite{Bertrand2011a},\cite{Cattivelli2010a}, \cite{Arablouei2014}. Here, we study the performance of PDRLS algorithm considering both sequential and stochastic partial-diffusion schemes under noisy information exchange.
\subsection{Assumptions}
Several simplifying assumptions have been traditionally adopted in the literature to gain insight into the performance of such adaptive algorithms. To proceed with the analysis we shall therefore introduce similar assumptions to what has been used before in the literature, and use them to derive useful performance measures.
\begin{assumption}
In order to make the analysis tractable, we introduce the following assumptions on statistical properties of the measurement data and noise signals.
\label{assu}
\end{assumption}
\begin{enumerate}
	\item The regression data $\boldsymbol{u}_{k,i}$ are temporally white and spatially independent random variables with zero mean and covariance matrix $\boldsymbol{R}_{u,k}\triangleq\E{\boldsymbol{u}^{\ast}_{k,i}\boldsymbol{u}_{k,i}}\geq0$.
	\item The noise signal $\boldsymbol{v}_k(i)$ and $\boldsymbol{v}^{\left(\psi\right)}_{k,i}$ are temporally white and spatially independent random variable with zero mean and co(variance) $\sigma^{2}_{v,k}$ and $R^{\left(\psi\right)}_{v,k}$, respectively. In addition, The quantities $\left\{R^{\left(\psi\right)}_{v,lk}\right\}$ are all zero if $l\in{\mathcal{N}_k}$ or when $l=k$.
	\item The regression data $\left\{\boldsymbol{u}_{m,i_{1}}\right\}$, the model noise signals $\boldsymbol{v}_n\left(i_{2}\right)$, and the link noise signals $\boldsymbol{v}^{\left(\psi\right)}_{l_{1}k_{1},j_{1}}$  are mutually independent random variables for all indexes $\left\{i_{1},i_{2},j_{1},k_{1},l_{1},m,n\right\}$.
	\item In order to make the performance analysis tractable, we introduce the following ergodicity assumption.
	For sufficiently large $i$, at any node $k$, we can replace $\boldsymbol{P}_{k,i}$ and $\boldsymbol{P}^{-1}_{k,i}$ with their expected values, $\E{\boldsymbol{P}_{k,i}}$ and $\E{\boldsymbol{P}^{-1}_{k,i}}$, respectively.
	\item For a sufficiently large $i$, at any node $k$, we have $\E{\boldsymbol{P}_{k,i}}=\E{\boldsymbol{P}^{-1}_{k,i}}^{-1}$
\end{enumerate}

Assumption 4 is a common assumption in the analysis of performance of RLS-type algorithm (see for example, \cite{sayed2011adaptive}, pp.318-319) and the results from assuming that the regressor vector of each node is an ergodic process, so that the time average of the node's rank-one instantaneous regressor covariance matrix, $\boldsymbol{u}^{\ast}_{k,i}\boldsymbol{u}_{k,i}$, over a sufficiently long time range can be replaced with the ensemble average (expected value). Assumption 5 is a good approximation when $\lambda$ is close to unity and the condition number of $\boldsymbol{R}_{u,k}$ is not very large \cite{Bertrand2011a}, \cite{Cattivelli2008}, \cite{Arablouei2014}, \cite{sayed2011adaptive}.

\subsection{Network Update Equation}
Our objective is to examine whether, and how fast, the weight estimates $\boldsymbol{w}_{k,i}$ from the distributed implementation \eqref{eq10}, \eqref{eq11}, and \eqref{eq20} converge towards the solution $\boldsymbol{w}^{o}$ \eqref{eq5}. To do so, we introduce the M×1 error vectors:
\begin{equation}
\boldsymbol{\tilde{\psi}}_{k,i}\triangleq \boldsymbol{w}^{o}-\boldsymbol{\psi}_{k,i}
\label{eq24}
\end{equation}
\begin{equation}
\boldsymbol{\tilde{w}}_{k,i}\triangleq \boldsymbol{w}^{o}-\boldsymbol{w}_{k,i}
\label{eq25}
\end{equation}

Furthermore, denote the network intermediate estimate-error and estimate-error vector as
\begin{equation}
\boldsymbol{\tilde{\psi}}_i\triangleq \mathrm{col} \left\{\boldsymbol{\tilde{\psi}}_{1,i},\ldots,\boldsymbol{\tilde{\psi}}_{N,i}\right\}
\label{eq26}
\end{equation}
\begin{equation}
\boldsymbol{\tilde{w}}_i\triangleq \mathrm{col} \left\{\boldsymbol{\tilde{w}}_{1,i},\ldots,\boldsymbol{\tilde{w}}_{N,i}\right\}
\label{eq27}
\end{equation}
Also, collecting the noise signal \eqref{eq21} and its covariances from across the network into $N\times1$ block vectors and $N\times N$ block diagonal matrices as follows:
\begin{equation}
\boldsymbol{v}^{\left(\psi\right)}_{i}\triangleq \mathrm{col} \left\{\boldsymbol{v}^{\left(\psi\right)}_{1,i},\ldots,\boldsymbol{v}^{\left(\psi\right)}_{N,i}\right\}
\label{eq28}
\end{equation}
\begin{equation}
\boldsymbol{R}^{\left(\psi\right)}_{v}\triangleq \mathrm{col}\left\{\boldsymbol{R}^{\left(\psi\right)}_{v,1},\ldots,\boldsymbol{R}^{\left(\psi\right)}_{v,N}\right\}
\label{eq29}
\end{equation}
Using the data model \eqref{eq1} and subtracting $\boldsymbol{w}^{o}$ from both sides of the relation \eqref{eq12} we obtain
\begin{equation}
	\boldsymbol{\tilde{\psi}}_{k,i}=\boldsymbol{\tilde{w}}_{k,i-1}-\boldsymbol{P}_{k,i}\boldsymbol{u}^{*}_{k,i}\left[\boldsymbol{u}_{k,i}\boldsymbol{\tilde{w}}_{k,i-1}+\boldsymbol{v}_{k}\left(i\right)\right]
	\label{eq30}
\end{equation}
which can be written as
\begin{eqnarray}
\boldsymbol{\tilde{\psi}}_{k,i}=\boldsymbol{P}_{k,i}\boldsymbol{P}^{-1}_{k,i}\boldsymbol{\tilde{w}}_{k,i-1}-\boldsymbol{P}_{k,i}\boldsymbol{u}^{*}_{k,i}\boldsymbol{u}_{k,i}\boldsymbol{\tilde{w}}_{k,i-1}\nonumber\\
	-\boldsymbol{P}_{k,i}\boldsymbol{u}^{*}_{k,i}\boldsymbol{v}_{k}\left(i\right)
	\label{eq31}
\end{eqnarray}
rewriting \eqref{eq12} as
\begin{equation}
	\boldsymbol{P}^{-1}_{k,i}=\lambda\boldsymbol{P}^{-1}_{k,i-1}+\boldsymbol{u}^{*}_{k,i}\boldsymbol{u}_{k,i}
	\label{eq32}
\end{equation}
Substituting \eqref{eq32} into the first term on RHS of \eqref{eq31} yields
\begin{equation}
	\boldsymbol{\tilde{\psi}}_{k,i}=\lambda\boldsymbol{P}_{k,i}\boldsymbol{P}^{-1}_{k,i-1}\boldsymbol{\tilde{w}}_{k,i-1}-\boldsymbol{P}_{k,i}\boldsymbol{u}^{*}_{k,i}\boldsymbol{v}_{k}\left(i\right)
	\label{eq33}
\end{equation}

We are interested in the steady-state behavior of the matrix $\boldsymbol{P}_{k,i}$. As $i\rightarrow\infty$, and $0\ll\lambda\leq1$, the steady-state mean value of $\boldsymbol{P}^{-1}_{k,i}$ is given by
\begin{align}
	\lim_{i\to\infty} \E{\boldsymbol{P}_{k,i}}&=	\lim_{i\to\infty} \E{\sum^{i}_{j=1}\lambda^{i-j}\boldsymbol{u}^{*}_{k,i}\boldsymbol{u}_{k,i}}\nonumber\\
	&=\lim_{i\to\infty}\sum^{i}_{j=1}\lambda^{i-j} \E{\boldsymbol{u}^{*}_{k,i}\boldsymbol{u}_{k,i}} \nonumber\\
	&=\left(\lim_{i\to\infty}\sum^{i}_{j=1}\lambda^{i-j}\right)\boldsymbol{R}_{u,k}\nonumber\\
	&=\frac{1}{1-\lambda}\boldsymbol{R}_{u,k}
	\label{eq34}
\end{align}

Using Assumptions (4) and (5) and relation \eqref{eq31}, we have for large enough $i$
\begin{align} \label{eq35}
	\boldsymbol{P}_{k,i}\boldsymbol{P}^{-1}_{k,i-1}& \approx \E{\boldsymbol{P}_{k,i}} \E{\boldsymbol{P}^{-1}_{k,i-1}}\nonumber \\
	& \approx \E{\boldsymbol{P}^{-1}_{k,i}}^{-1} \E{\boldsymbol{P}^{-1}_{k,i-1}}\nonumber \\
	& \approx \boldsymbol{I}_{M}	
\end{align}
and
\begin{equation}
	\boldsymbol{P}_{k,i}\approx \E{\boldsymbol{P}_{k,i}}\approx \E{\boldsymbol{P}^{-1}_{k,i}}^{-1}
	\approx\left(1-\lambda\right)\boldsymbol{R}^{-1}_{u,k}
	\label{eq36}
\end{equation}

Consequently for a sufficiently large $i$, \eqref{eq33} can be approximated by
\begin{equation}
	\boldsymbol{\tilde{\psi}}_{k,i}=\lambda\boldsymbol{\tilde{w}}_{k,i}-\left(1-\lambda\right)\boldsymbol{R}^{-1}_{u,k}\boldsymbol{u}^{*}_{k,i}v_{k}\left(i\right)
	\label{eq37}
\end{equation}
On the other hand, subtracting both sides of \eqref{eq20} from $w^{o}$ gives
\begin{eqnarray}
	\boldsymbol{\tilde{w}}_{k,i}=\left(\boldsymbol{I}_{M}-\sum_{l\in{\mathcal{N}_k}\backslash\left\{k\right\}}a_{lk}\boldsymbol{\mathcal{K}}_{l,i}\right)\boldsymbol{\tilde{\psi}}_{k,i}\nonumber\\
	+\sum_{l\in{\mathcal{N}_k}\backslash\left\{k\right\}}a_{lk}\boldsymbol{\mathcal{K}}_{l,i}\boldsymbol{\tilde{\psi}}_{l,i}-\boldsymbol{v}^{\left(\psi\right)}_{k,i}
	\label{eq38}
\end{eqnarray}
To describe these relations more compactly, we collect the information from across the network into block vectors and matrices. Using \eqref{eq26}-\eqref{eq28}, leads to
\begin{equation}
	\boldsymbol{\tilde{\psi}}_{i}=\lambda\boldsymbol{\tilde{w}}_{i-1}-\boldsymbol{\Gamma}\boldsymbol{s}_{i}
	\label{eq39}
\end{equation}
\begin{equation}
	\boldsymbol{\tilde{w}}_{i}=\boldsymbol{\mathcal{B}}_{i}\boldsymbol{\tilde{\psi}}_{i-1}-\boldsymbol{v}^{\left(\psi\right)}_{i}
	\label{eq40}
\end{equation}
where
\begin{equation}
	\boldsymbol{\Gamma}\triangleq \left(1-\lambda\right)\mathrm{diag}\left\{\boldsymbol{R}^{-1}_{u,1},\ldots,\boldsymbol{R}^{-1}_{u,N}\right\}
	\label{eq41}
\end{equation}
\begin{equation}
	\boldsymbol{s}_i\triangleq \left\{\boldsymbol{u}^{*}_{1,i}\boldsymbol{v}_{1}(i),\ldots,\boldsymbol{u}^{*}_{N,i}\boldsymbol{v}_{N}(i)\right\}
	\label{eq42}
\end{equation}
\begin{equation}
%\left.\begin{aligned}
\boldsymbol{\mathcal{B}}_{i}=
\begin{bmatrix}
\boldsymbol{B}_{1,1,i} & \cdots & \boldsymbol{B}_{1,N,i} \\
\vdots & \ddots & \vdots \\
\boldsymbol{B}_{N,1,i} & \cdots & \boldsymbol{B}_{N,N,i}
\end{bmatrix}
%\end{aligned}
\label{eq43}
\end{equation}
where
\begin{equation}
\boldsymbol{B}_{p,q,i}=\begin{cases}
\boldsymbol{I}_M-\sum_{l\in{\mathcal{N}_p}\backslash\left\{p\right\}}a_{lp}\boldsymbol{\mathcal{K}}_{l,i} & \text{if}\, p=q \\
a_{qp}\boldsymbol{\mathcal{K}}_{q,i} & \text{if}\, q\in{\mathcal{N}_p}\backslash\left\{p\right\} \\
\boldsymbol{O}_M & \text{otherwise}
\end{cases}
\label{eq44}
\end{equation}
and $\boldsymbol{O}_{M}$ is the $M\times M$ zero matrix.
So, the network weight error vector, $\boldsymbol{\tilde{w}}_{i}$, ends up evolving according to the following stochastic recursion:
\begin{equation}
	\boldsymbol{\tilde{w}}_{i}=\lambda\boldsymbol{\mathcal{B}}_{i}\boldsymbol{\tilde{w}}_{i-1}-\boldsymbol{\mathcal{B}}_{i}\boldsymbol{\Gamma}\boldsymbol{s}_{i}-\boldsymbol{v}^{\left(\psi\right)}_{i}
	\label{eq45}
\end{equation}
\subsection{Convergence in Mean}
Taking expectation of both sides of \eqref{eq45} under \textit{Remark} and \textit{Assumptions} \textit{1} and \textit{2}, we find that the mean error vector evolves according to the following recursion:
\begin{equation}
	\E{\boldsymbol{\tilde{w}}}_i=\lambda\boldsymbol{Q}\E{\boldsymbol{\tilde{w}}_{i-1}}
	\label{eq46}
\end{equation}
where
\begin{equation}
	\boldsymbol{Q}=\E{\boldsymbol{\mathcal{B}}_{i}}
	\label{eq47}
\end{equation}
Like \cite{Arablouei2014a}, $\boldsymbol{Q}$ can be obtained for both stochastic and sequential partial-diffusion using the definition of $\boldsymbol{\mathcal{B}}_{i}$. What is most noteworthy here is to find the value of each $\boldsymbol{Q}$ entry after applying expectation operator. Therefore, we can write
\begin{align}
\E{\boldsymbol{B}_{p,q,i}}&=\begin{cases}
\left(1-\rho+\rho a_{r,lp}\right)\boldsymbol{I}_{M} & \text{if}\, p=q \\
\rho a_{r,qp}\boldsymbol{I}_{M} & \text{if}\, q\in{\mathcal{N}_p}\backslash\left\{p\right\} \\
\boldsymbol{O}_M & \text{otherwise}
\end{cases}
\label{eq48}
\end{align}
All the entries of $\boldsymbol{Q}$ are real and non-negative and all the rows of $\boldsymbol{Q}$ add up to unity. This property can be established for both stochastic and sequential partial-diffusion schemes and for any value of $L$ \cite{Arablouei2014a}. This implies that $\boldsymbol{Q}$ is a right stochastic matrix. In view of the fact that the eigenvalue of a stochastic matrix with the largest absolute value (spectral radius) is equal to one \cite{meyer2000matrix,Al-Naffouri2003}, the spectral radius of the matrix $\lambda\boldsymbol{Q}$ is equal to $\lambda$. Thus, for $0\ll\lambda\leq1$, every element of $\E{\boldsymbol{\tilde{w}}_{i}}$ converges to zero as $i\rightarrow\infty$, and the estimator is asymptotically unbiased and convergent in mean. Note that this is not in fact the necessary and sufficient condition for convergence of $\E{\boldsymbol{\tilde{w}}_{i}}$ as \eqref{eq46} has been obtained under an independence assumption which is not true in general.

\subsection{Mean-Square Stability}
We now study the mean-square performance of PDRLS under imperfect information exchange. We will also derive closed-form expressions that characterize the network performance. Doing so, we resort to the energy conservation analysis of \cite{Arablouei2014}, \cite{Sayed2003}, \cite{Al-Naffouri2003}, \cite{yousef2001unified}. The details are as follows.

Equating the squared weighted Euclidean norm of both sides of \eqref{eq45} and applying the expectation operator together with using \textit{Remark} and \textit{Assumptions} \textit{1} and \textit{2} yield the following weighted variance relation:
\begin{align}
\E{\left\|\boldsymbol{\tilde{w}}_{i}\right\|^{2}_{\Sigma}} & = \E{\boldsymbol{\tilde{w}}^{*}_{i-1}\lambda^{2}\boldsymbol{\mathcal{B}}^{T}_{i} \boldsymbol{\Sigma} \boldsymbol{\mathcal{B}}_{i}\boldsymbol{\tilde{w}}_{i-1}}\nonumber\\
&+\E{\boldsymbol{s}^{*}_{i}\boldsymbol{\Gamma}\boldsymbol{\mathcal{B}}^{T}_{i}\boldsymbol{\Sigma}\boldsymbol{\mathcal{B}}_{i}\boldsymbol{\Gamma}\boldsymbol{s}_{i}}
+\E{\boldsymbol{v}^{{*}\left(\psi\right)}_{i}\boldsymbol{\Sigma}\boldsymbol{v}^{\left(\psi\right)}_{i}}
\label{eq49}
\end{align}
where $\boldsymbol{\Sigma}$ is an arbitrary symmetric nonnegative-definite matrix. Let us evaluate each of the expectations on the right-hand side. The first expectation is given by
\begin{align}
	\E{\boldsymbol{\tilde{w}}^{*}_{i-1}\lambda^{2}\boldsymbol{\mathcal{B}}^{T}_{i}\boldsymbol{\Sigma} \boldsymbol{\mathcal{B}}_{i}\boldsymbol{\tilde{w}}_{i-1}} 
	& \nonumber\\
	& =\E{\E{\boldsymbol{\tilde{w}}^{*}_{i-1}\lambda^{2}\boldsymbol{\mathcal{B}}^{T}_{i}\boldsymbol{\Sigma} 	\boldsymbol{\mathcal{B}}_{i}\boldsymbol{\tilde{w}}_{i-1}|\boldsymbol{\tilde{w}}_{i-1}}}\nonumber\\
	& =\E{\boldsymbol{\tilde{w}}^{*}_{i-1}}\E{\lambda^{2}\boldsymbol{\mathcal{B}}^{T}_{i}\boldsymbol{\Sigma} \boldsymbol{\mathcal{B}}_{i}}\nonumber\\
	& \triangleq\E{\boldsymbol{\tilde{w}}^{*}_{i-1}\boldsymbol{\Sigma}'\boldsymbol{\tilde{w}}_{i-1}}\nonumber\\
& = \E{\left\|\boldsymbol{\tilde{w}}_{i}\right\|^{2}_{\boldsymbol{\Sigma'}}}
\label{eq50}
\end{align}
where we introduce the nonnegative-definite weighting matrix
\begin{equation}
	\boldsymbol{\Sigma'}=\E{\lambda^{2}\boldsymbol{\mathcal{B}}^{T}_{i}\boldsymbol{\Sigma} \boldsymbol{\mathcal{B}}_{i}}
	\label{eq51}
\end{equation}
Since $\boldsymbol{\tilde{w}}_{i-1}$ is independent of $\boldsymbol{\Sigma'}$, we have
\begin{equation}
	\E{\left\|\boldsymbol{\tilde{w}}_{i-1}\right\|^{2}_{\boldsymbol{\Sigma'}}}=\E{\left\|\boldsymbol{\tilde{w}}_{i-1}\right\|^{2}_{\E{\boldsymbol{\Sigma'}}}}
	\label{eq52}
\end{equation}
It is convenient to introduce the alternative notation $\left\|\boldsymbol{x}\right\|^{2}_{\boldsymbol{\sigma}}$ to refer to the weighted square quantity $\left\|\boldsymbol{x}\right\|^{2}_{\boldsymbol{\Sigma}}$, where $\boldsymbol{\sigma}=\vect{\boldsymbol{\Sigma}}$. We shall use these two notations interchangeably.

Using the following equalities for arbitrary matrices $\left\{\boldsymbol{U},\boldsymbol{W},\boldsymbol{\Sigma},\boldsymbol{Z}\right\}$ of compatible dimensions:
\begin{eqnarray}
\label{eq53}
\left(\boldsymbol{U}\otimes \boldsymbol{W}\right)\left(\boldsymbol{\Sigma}\otimes \boldsymbol{Z}\right) & = & \boldsymbol{U}\boldsymbol{\Sigma}\otimes \boldsymbol{W}\boldsymbol{Z} \\
\vect{\boldsymbol{U}\boldsymbol{\Sigma}\boldsymbol{W}} & = &\left(\boldsymbol{W}^{T}\otimes \boldsymbol{U}\right)\vect{\boldsymbol{\Sigma}} \label{eq54} \\
\mathrm{Tr}\left(\boldsymbol{\Sigma} \boldsymbol{W}\right) & = & \left[\vect{\boldsymbol{W}^{T}}\right]^{T}\vect{\boldsymbol{\Sigma}}
\label{eq55}
\end{eqnarray}
we have
\begin{equation}
	\boldsymbol{\sigma'}=\E{\lambda^{2}\boldsymbol{\mathcal{B}}^{T}_{i}\boldsymbol{\Sigma}\boldsymbol{\mathcal{B}}_{i}}=\boldsymbol{\mathcal{F}}\boldsymbol{\sigma}
	\label{eq56}
\end{equation}
where
\begin{equation}
	\boldsymbol{\mathcal{F}}=\lambda^{2}\boldsymbol{\Phi}
	\label{eq57}
\end{equation}
\begin{equation}
	\boldsymbol{\Phi}=\mathbb{E}\boldsymbol{\mathcal{B}}^{T}_{i}\otimes\boldsymbol{\mathcal{B}}^{T}_{i}
	\label{eq58}
\end{equation}
Here, the efforts to find \eqref{eq58} in \cite{Arablouei2014a} can be extended. Therefore, $\boldsymbol{\Phi}$ can be established in general form. Again, what is most important here is to discuss about the probability of transmitted entries of nodes. This would be helpful in finding some expectations, $\E{\kappa_{t,p,i}\boldsymbol{\mathcal{K}}_{q,i}}$, we are faced with.

At a given iteration, the probability of transmitting two different entries of single node is \cite{Arablouei2014a}
\begin{equation}
\rho\left(\frac{L-1}{M-1}\right)
\label{eq59}
\end{equation}
In the stochastic partial-diffusion scheme, at a given iteration, the probability of transmitting two entries from two different nodes is $\rho^2$. Thus, we have \cite{Arablouei2014a}
\begin{align}
\E{\kappa_{t,p,i}\boldsymbol{\mathcal{K}}_{q,i}}\nonumber\\
&=\begin{cases}
\rho\left(\frac{L-1}{M-1}\right)\boldsymbol{I}_{M}+\rho\left(\frac{M-L}{M-1}\right)\boldsymbol{\mathcal{J}}^{t,t}_{M} & \text{if}\, p=q \\
\rho^{2} \boldsymbol{I}_{M} & \text{if}\, p\neq q
\end{cases}
\label{eq60}
\end{align}
where $\boldsymbol{\mathcal{J}}^{t,t}_{M}$ is an $M\times M$ single-entry matrix with one at its $\left(t,t\right)$th entry and zeros elsewhere, for $p=1,\ldots,N$, $q=1,\ldots,N$, and $t=1,\ldots,M$.
Furthermore, in sequential partial-diffusion and under same entry selection pattern, at a given iteration, we have \cite{Arablouei2014a}
\begin{equation}
\E{\kappa_{t,p,i}\boldsymbol{\mathcal{K}}_{q,i}}=\rho\left(\frac{L-1}{M-1}\right)\boldsymbol{I}_{M}+\rho\left(\frac{M-L}{M-1}\right)\boldsymbol{\mathcal{J}}^{t,t}_{M}
\label{eq61}
\end{equation}
As shown in \cite{Arablouei2014a}, all the entries of $\boldsymbol{\Phi}$ are real and non-negative. moreover, all their columns sum up to one both for stochastic and sequential partial-diffusion scheme and for any value of $L$.

Second term on RHS of \eqref{eq49} 
\begin{eqnarray}
	\E{\boldsymbol{s}^{*}_{i}\boldsymbol{\Gamma}\boldsymbol{\mathcal{B}}^{T}_{i}\boldsymbol{\Sigma}\boldsymbol{\mathcal{B}}_{i}\boldsymbol{\Gamma}\boldsymbol{s}_{i}}=\mathrm{vec}^T\left\{\boldsymbol{\mathcal{G}}\right\}\boldsymbol{\Phi}\boldsymbol{\sigma}
	\label{eq62}
%\E{Tr\left\{\Sigma\boldsymbol{\mathcal{A}}_{i}\boldsymbol{\Gamma}\boldsymbol{s}_{i}\boldsymbol{s}^{*}_{i}\boldsymbol{\Gamma}\boldsymbol{\mathcal{A}}^{T}_{i}\right\}\nonumber\\
	%&=vec^{T}\left\{\mathbb{E}\left(\boldsymbol{\mathcal{A}}_{i}\boldsymbol{\Gamma}\boldsymbol{s}_{i}\boldsymbol{s}^{*}_{i}\boldsymbol{\Gamma}\boldsymbol{\mathcal{A}}^{T}_{i}\right)\right\}\nonumber\\
	%&=\left(\boldsymbol{\Phi}^{T}vec\left\{\mathbb{E}\left(\boldsymbol{\Gamma}\boldsymbol{s}_{i}\boldsymbol{s}^{*}_{i}\boldsymbol{\Gamma}\right)\right\}\right)^{T}\boldsymbol{\sigma}\nonumber\\
\end{eqnarray}
where
\begin{equation}
	\boldsymbol{\mathcal{G}}=\boldsymbol{\Gamma}\E{\boldsymbol{s}_{i}\boldsymbol{s}^{*}_{i}}\boldsymbol{\Gamma}
	\label{eq63}
\end{equation}
which, in view of \textit{Assumptions}, can be expressed as
\begin{equation}
	\boldsymbol{\mathcal{G}}=\left(1-\lambda\right)^{2}\left\{\sigma^{2}_{v,1}\boldsymbol{R}^{-1}_{u,1},\ldots,\sigma^{2}_{v,N}\boldsymbol{R}^{-1}_{u,N}\right\}
	\label{eq64}
\end{equation}

Last term on RHS of \eqref{eq49}:
\begin{eqnarray}
\E{\boldsymbol{v}^{{*}\left(\psi\right)}_{i}\boldsymbol{\Sigma}\boldsymbol{v}^{\left(\psi\right)}_{i}}=\E{\mathrm{Tr}\left\{\boldsymbol{\Sigma}\boldsymbol{v}^{\left(\psi\right)}_{i}\boldsymbol{v}^{{*}\left(\psi\right)}_{i}\right\}}\nonumber\\
	=\mathrm{vec}^T\left(\boldsymbol{R}^{\left(\psi\right)}_{v}\right)\boldsymbol{\sigma}
	\label{eq65}
\end{eqnarray}

The variance relation becomes
\begin{eqnarray}	\E{\left\|\boldsymbol{\tilde{w}}_{i}\right\|^{2}_{\sigma}}=\E{\left\|\boldsymbol{\tilde{w}}_{i-1}\right\|^{2}_{\lambda^{2}\Phi\sigma}}+\mathrm{vec}^T\left\{\boldsymbol{\mathcal{G}}\right\}\boldsymbol{\Phi}\boldsymbol{\sigma}\nonumber\\
	+\mathrm{vec}^T\left(\boldsymbol{R}^{\left(\psi\right)}_{v}\right)\boldsymbol{\sigma}
	\label{eq66}
\end{eqnarray}

A recursion of type \eqref{eq66} is stable and convergent if the matrix $\lambda^{2}\boldsymbol{\Phi}$ is stable \cite{sayed2011adaptive}. The entries of $\boldsymbol{\Phi}$ are all real-valued and non-negative. In addition, all the columns of $\boldsymbol{\Phi}$ add up to unity. As a result, $\boldsymbol{\Phi}$ is left-stochastic and has unit spectral radius. Hence, the spectral radius of the matrix $\lambda^{2}\boldsymbol{\Phi}$ is equal to $\lambda^{2}$ that is smaller than one. Therefore, the mean-square stability condition for PDRLS under noisy links is the same as that of PDRLS under noise-free links and the rate of this convergence is dependent on the value of $\lambda$. Again, this need not translate to be the necessary and sufficient condition for convergence of recursion of type \eqref{eq66} in actuality as \eqref{eq66} has been obtained under independence assumption which is not true in general. Therefore, the mean-square convergence of PDRLS under noisy links is an open question. 

\subsection{Mean-Square Performance}
At steady state, when $i\rightarrow\infty$, \eqref{eq66} can be written as
\begin{align} \label{eq67}
	\lim_{i\to\infty} \E{\left\|\tilde{\boldsymbol{w}}_{i}\right\|^{2}_{\left(\boldsymbol{I}_{N^{2}M^{2}}-\boldsymbol{\mathcal{F}}\right)\boldsymbol{\sigma}}}&= \nonumber \\
	&\mathrm{vec}^T\left\{\boldsymbol{\mathcal{G}}\right\}\boldsymbol{\Phi}\boldsymbol{\sigma}
	+\mathrm{vec}^T\left\{\boldsymbol{R}^{\left(\psi\right)}_{v}\right\}\boldsymbol{\sigma} \nonumber \\
\end{align}
Expression \eqref{eq66} is a very useful relation; it allows us to evaluate the network MSD through proper selection of the weighting vector $\boldsymbol{\sigma}$. The network MSD is defined as the average value:
\begin{equation}
\mathrm{MSD}^{\mathrm{network}}\triangleq\lim_{i\to\infty}\frac{1}{N}\sum_{k=1}^{N}\E{\left\|\boldsymbol{\tilde{w}}_{k,i}\right\|^{2}}
\label{eq68}
\end{equation}
which amounts to averaging the MSDs of the individual nodes. Therefore,
\begin{equation}
\mathrm{MSD}^{\mathrm{network}}=\lim_{i\to\infty}\frac{1}{N}\E{\left\|\boldsymbol{\tilde{w}}_{i}\right\|^{2}}=\lim_{i\to\infty}\E{\left\|\boldsymbol{\tilde{w}}_{i}\right\|^{2}_{1/N}}
\label{eq69}
\end{equation}
This means that in order to recover the network MSD from relation \eqref{eq66}, we should select the weighting vector $\sigma$ such that
\begin{equation}
\left(\boldsymbol{I}_{N^{2}M^{2}}-\boldsymbol{\mathcal{F}}\right)\sigma=\frac{1}{N}\vect{\boldsymbol{I}_{NM}}
\label{eq70}
\end{equation}
Solving for $\sigma$ and substituting back into \eqref{eq66} we arrive at the following expression for the network MSD
\begin{eqnarray}
\mathrm{MSD}^{\mathrm{network}}_{\mathrm{noisy}}&=\frac{1}{N}\left[\mathrm{vec}^T\left\{\boldsymbol{\mathcal{G}}\right\}\boldsymbol{\Phi}+\mathrm{vec}^T\left\{\boldsymbol{R}^{\left(\psi\right)}_{v}\right\}\right]\nonumber\\
&\left(\boldsymbol{I}_{N^{2}M^{2}}-\boldsymbol{\mathcal{F}}\right)^{-1}\vect{\boldsymbol{I}_{NM}}
\label{eq71}
\end{eqnarray}
In perfect information exchange, the last two terms of \eqref{eq66} don not appear, so we can conclude that the network MSD deteriorates as follows:
\begin{eqnarray}
\mathrm{MSD}^{\mathrm{network}}_{\mathrm{noisy}}&=&\mathrm{MSD}^{\mathrm{network}}_{\mathrm{ideal}}+\nonumber\\
&&\frac{1}{N}\mathrm{vec}^T\left\{\boldsymbol{R}^{\left(\psi\right)}_{v}\right\}\left(\boldsymbol{I}_{N^{2}M^{2}}-\boldsymbol{\mathcal{F}}\right)^{-1}\vect{\boldsymbol{I}_{NM}}\nonumber\\
\label{eq72}
\end{eqnarray}

\section{Simulation Results}
In order to illustrate the performance of each PDRLS strategy under imperfect information exchange, we consider an adaptive network with a random topology and $N=10$ where each node is, on average, connected to two other nodes. The measurements were generated according to model \eqref{eq1}, and regressors, $\boldsymbol{u}_{k,i}$, were chosen Gaussian i.i.d with randomly generated different diagonal covariance matrices, $\boldsymbol{R}_{u,k}$. The additive noises at nodes are zero mean Gaussian with variances $\boldsymbol{\sigma}^{2}_{v,k}$ and independent of the regression data. The traces of the covariance matrix regressors and the noise variances at all nodes, $Tr\left(\boldsymbol{R}_{u,k}\right)$ and $\boldsymbol{\sigma}^{2}_{v,k}$, are shown in Fig. 1. We also use white Gaussian link noise signals such that $R^{\left(\psi\right)}_{v,lk}=\sigma^{2}_{\psi,lk}I_{M}$. All link noise variances $\sigma^{2}_{\psi,lk}I_{M}$, are randomly generated and illustrated in Fig. 2. We assign the link number by the following procedure. We denote the link from node $l$ to node $k$ as $\ell_{l,k}$, where $l\neq k$. Then, we collect the links $\left\{\ell_{l,k}, l\in{\mathcal{N}_k}\backslash\left\{k\right\}\right\}$ in an ascending order of $l$ in the list $\mathcal{L}_{k}$ (which is a set with ordered elements) for each node. We concatenate $\left\{\mathcal{L}_{k}\right\}$ in an ascending order of $k$ to get the overall list $\mathcal{L}=\left\{\mathcal{L}_{1},\mathcal{L}_{2},\ldots,\mathcal{L}_{N}\right\}$. Eventually, the \textit{m}th link in the network is given by the \textit{m}th element in the list $\mathcal{L}$. It is noteworthy that we adopt the network MSD learning curves of all figures by averaging over 50 experiments and the unknown parameter $w^o$ of length $M=8$ is randomly generated. In Fig. 3, we illustrate the simulated time evolution of network MSD of the PDRLS algorithm under noisy information exchange using both sequential and stochastic partial-diffusion schemes for $\lambda=0.995$. Fig. 4, demonstrates a similar scenario as that in Fig. 3 when $\lambda=1$. To compare the steady-state performance of network MSD curves of PDRLS strategies, we examine the network MSD learning curve of PDRLS strategies again with ideal links under various numbers of entries communicated at each iteration, $L$, when $\lambda=1$ as illustrated in Fig. 5.
From the results above, we can make the following observations:
\begin{itemize}
\item The PDRLS algorithm delivers a tradeoff between communications cost and estimation performance under noise-free links \cite{Arablouei2014};
\item The MSD performance of PDRLS with noisy links is strictly come under the influence of forgetting factor, $\lambda$. So, a minimal change in $\lambda$ leads to a significant degradation on MSD performance;  
\item As can be seen, PDLMS algorithm with noisy links fails to converge for both  stochastic and sequential schemes, whereas it converges for the noise-free links case..
\item The more entries are communicated at each iteration, the more perturbed weight estimates are interred in the consultation phase. Therefore, the number of communicated entries has a marked effect on the MSD performance of PDRLS with noisy links.   
\end{itemize}

\begin{figure} [t]%htbp Fig1
\centering 
\includegraphics [width=8.5cm]{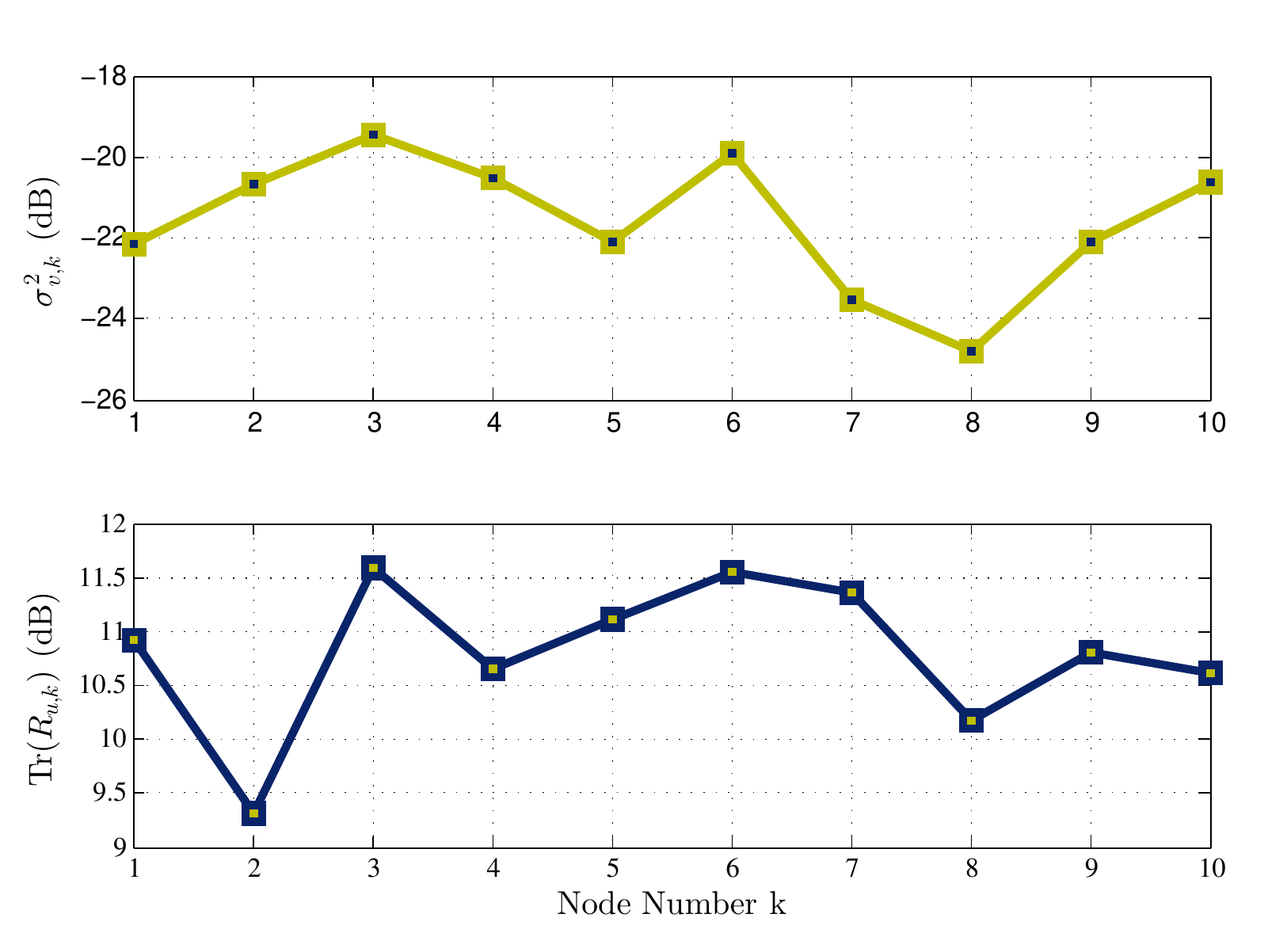} %   {figure-1.pdf}  %[width=8.5cm]   
\centering \caption{Variance of the noise (top) and Covariance matrix trace of the input signal (bottom) at each node.}
\label{fig:1}
\end{figure}
\begin{figure} [t]%htbp Fig1
\centering 
\includegraphics [width=8.5cm]{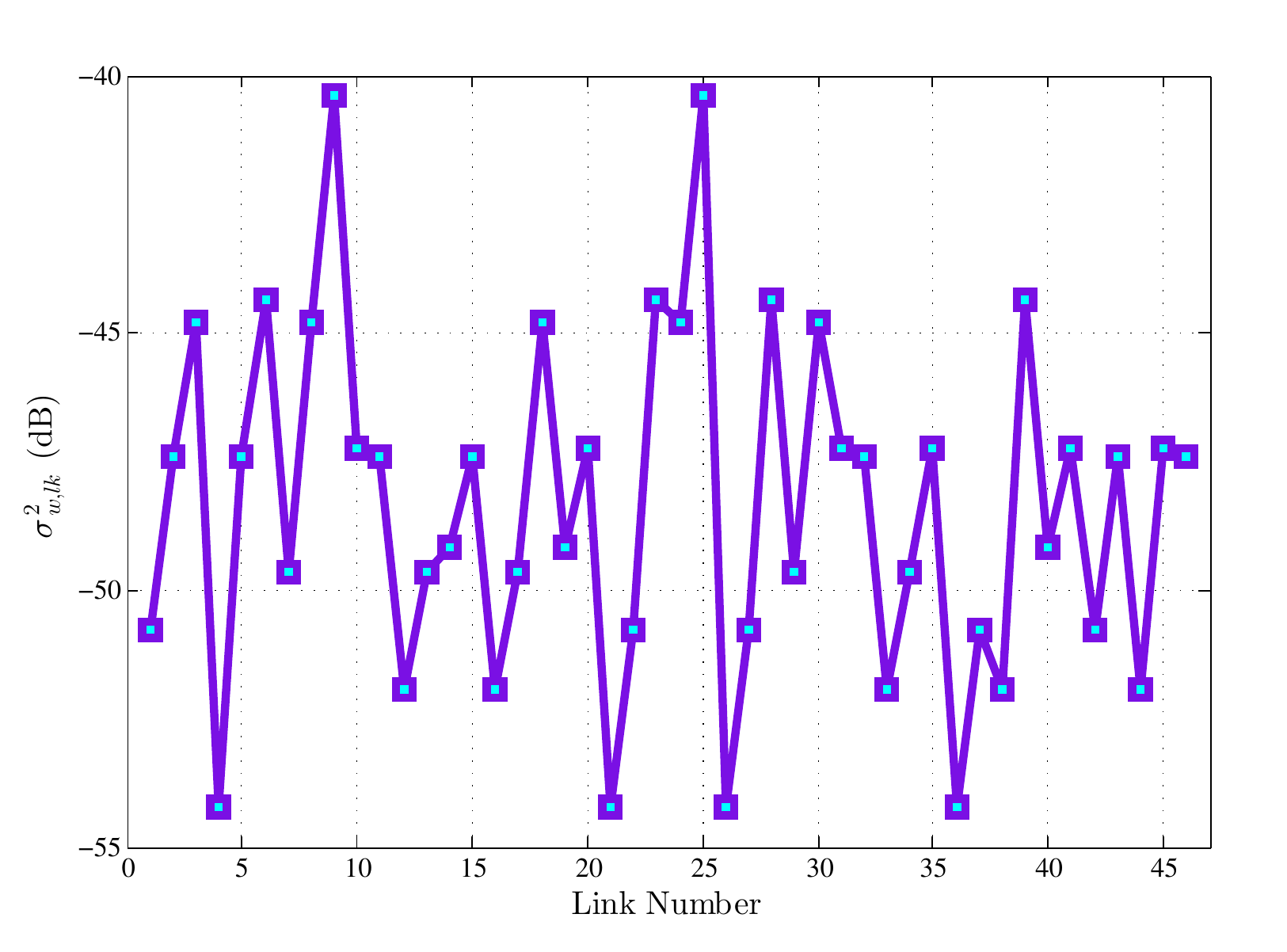} %   {figure-1.pdf}  %[width=8.5cm]   
\centering \caption{The variance profiles for various sources of link noises in dB, $\sigma^{2}_{\psi,lk}$.}
\label{fig:2}
\end{figure}
%%%%%%%%%%%%%%%%%%%%%%%%%%%%%%%%%%%%%%%%%%%%%%%
\begin{figure}  [t]% fig 10
  \centering
  \subfigure{\label{figtop:3}\includegraphics[width=8.5cm]{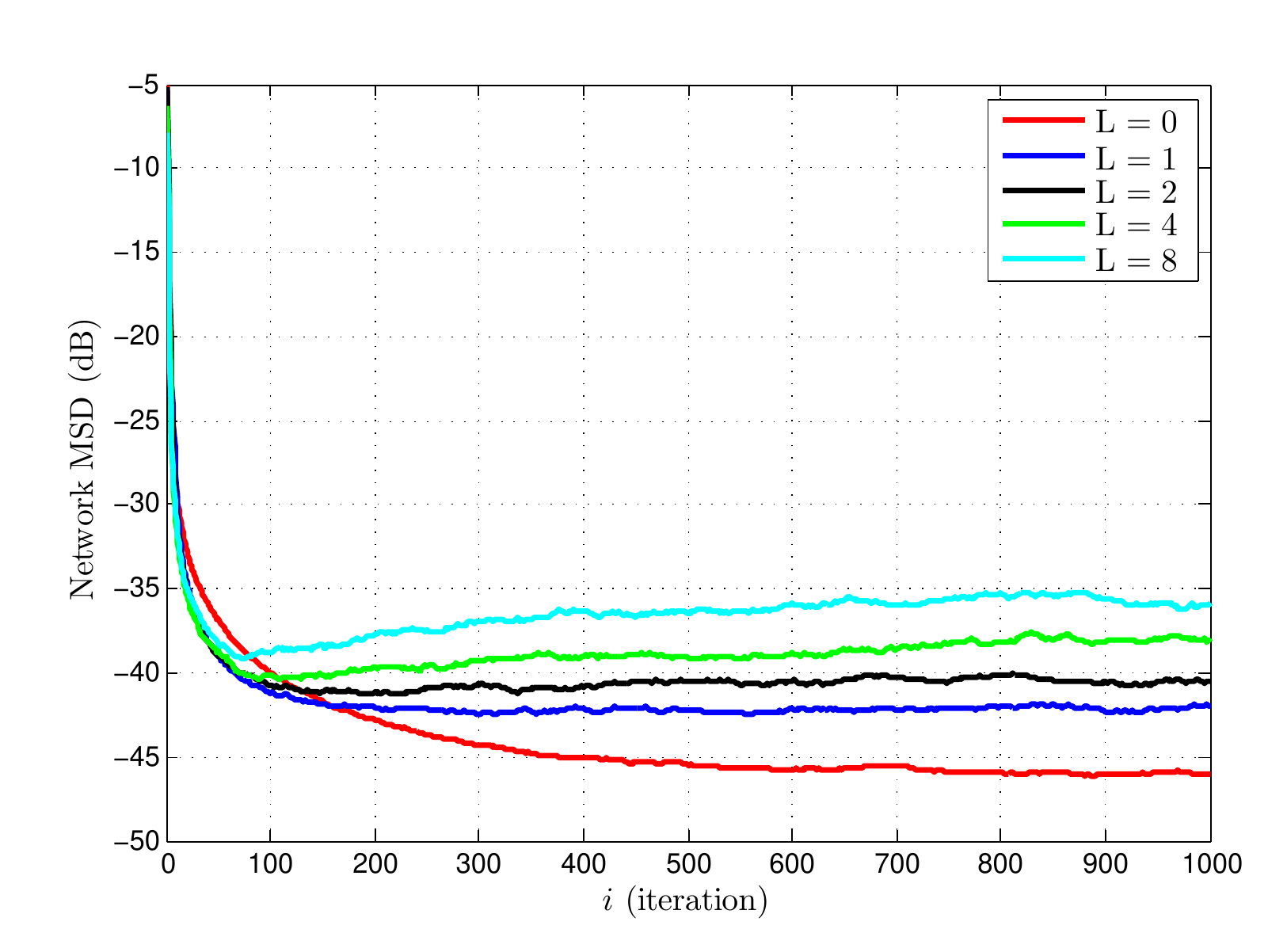}}  
  \subfigure{\label{figbottom:3}\includegraphics[width=8.5cm]{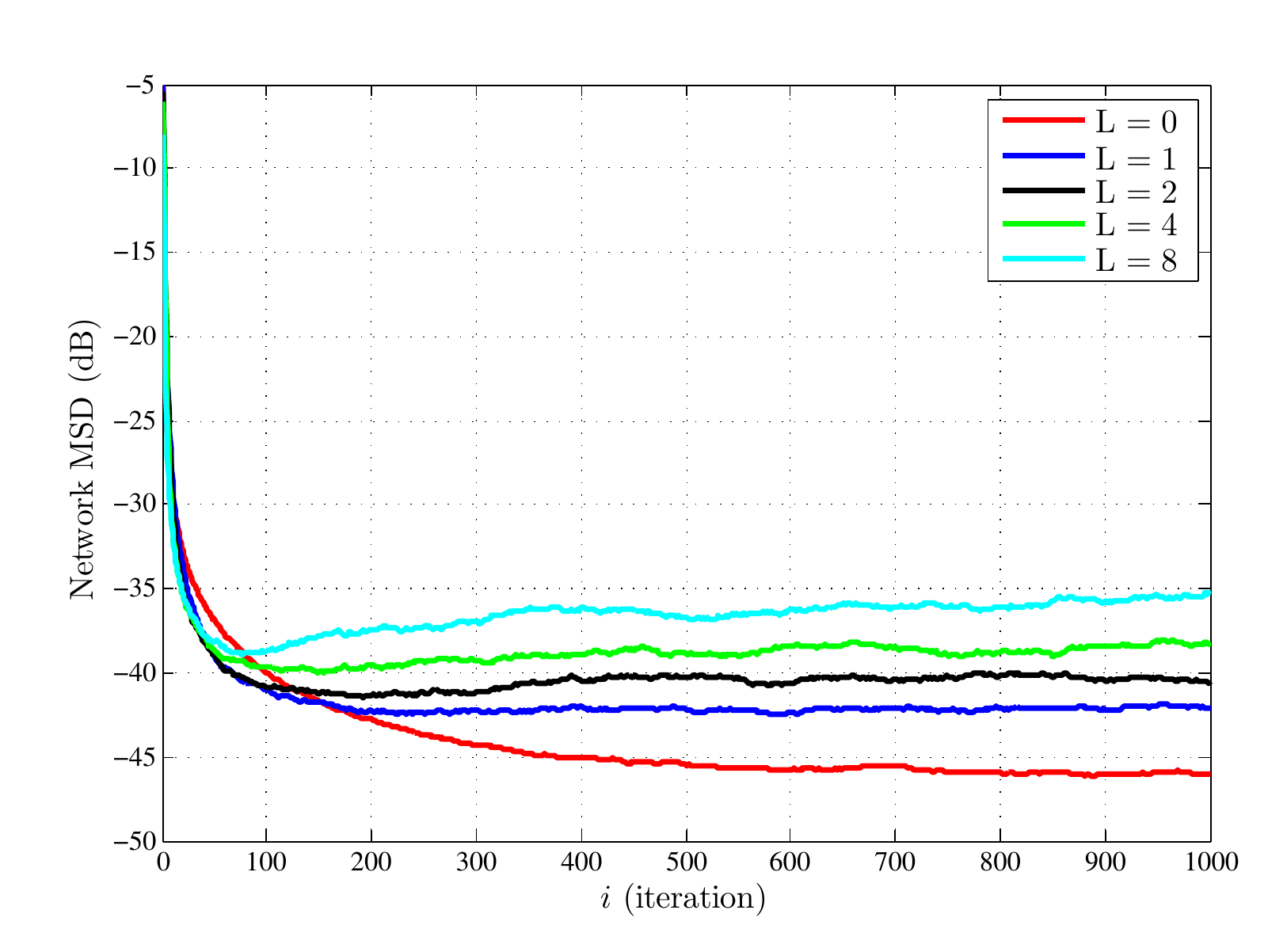}}   
  \caption{Simulated network MSD curves for partial diffusion RLS algorithms using sequential (top) and stochastic (bottom) with different number of entries communicated under noisy links when $\lambda=0.995$.}
 \end{figure}
%%%%%%%%%%%%%%%%%%%%%%%%%%%%%%%%%%%%%%%%%%%%%%%
%%%%%%%%%%%%%%%%%%%%%%%%%%%%%%%%%%%%%%%%%%%%%%%
\begin{figure}  [t]% fig 10
  \centering
  \subfigure{\label{figtop:4}\includegraphics[width=8.5cm]{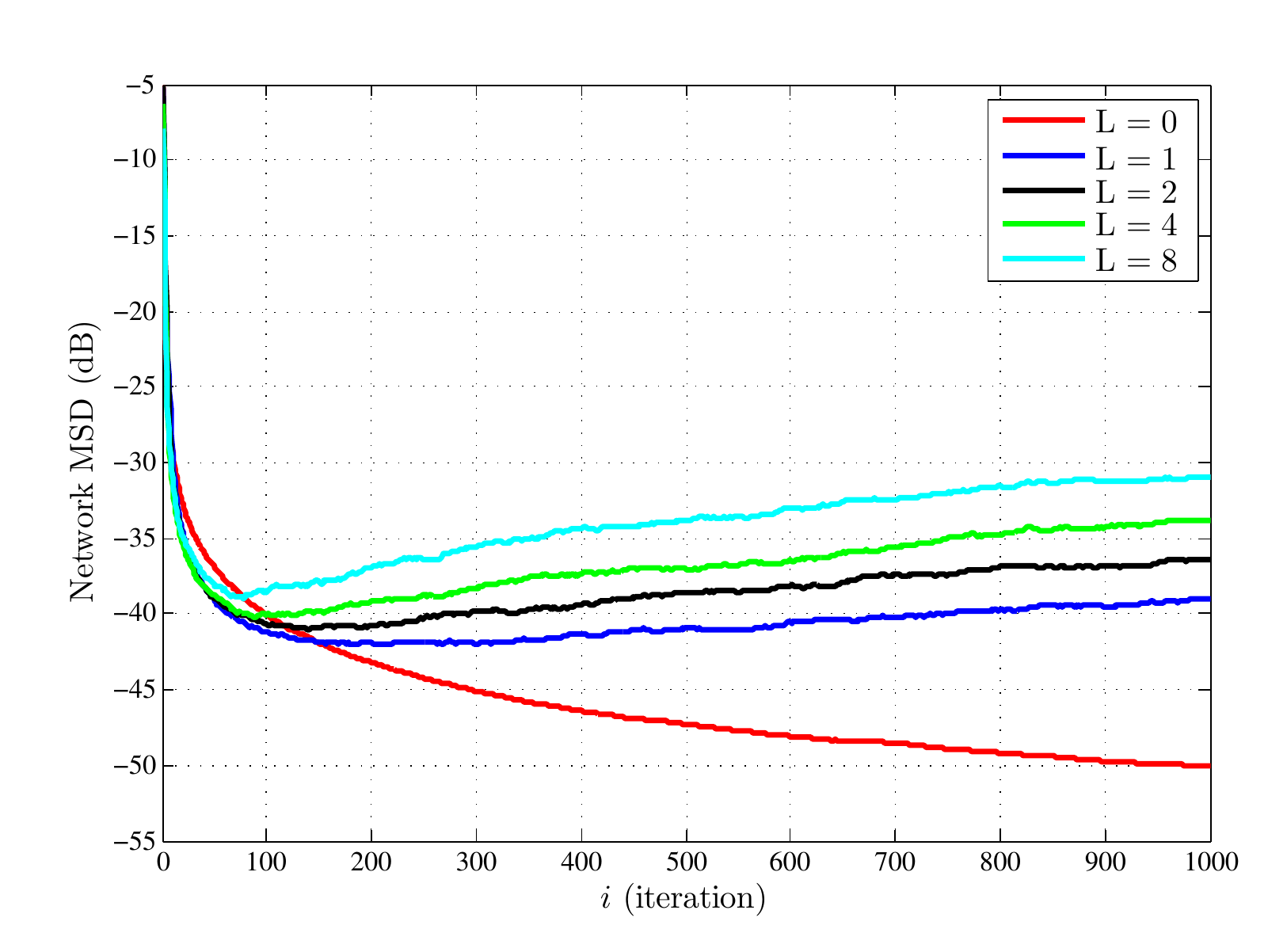}}  
  \subfigure{\label{figbottom:4}\includegraphics[width=8.5cm]{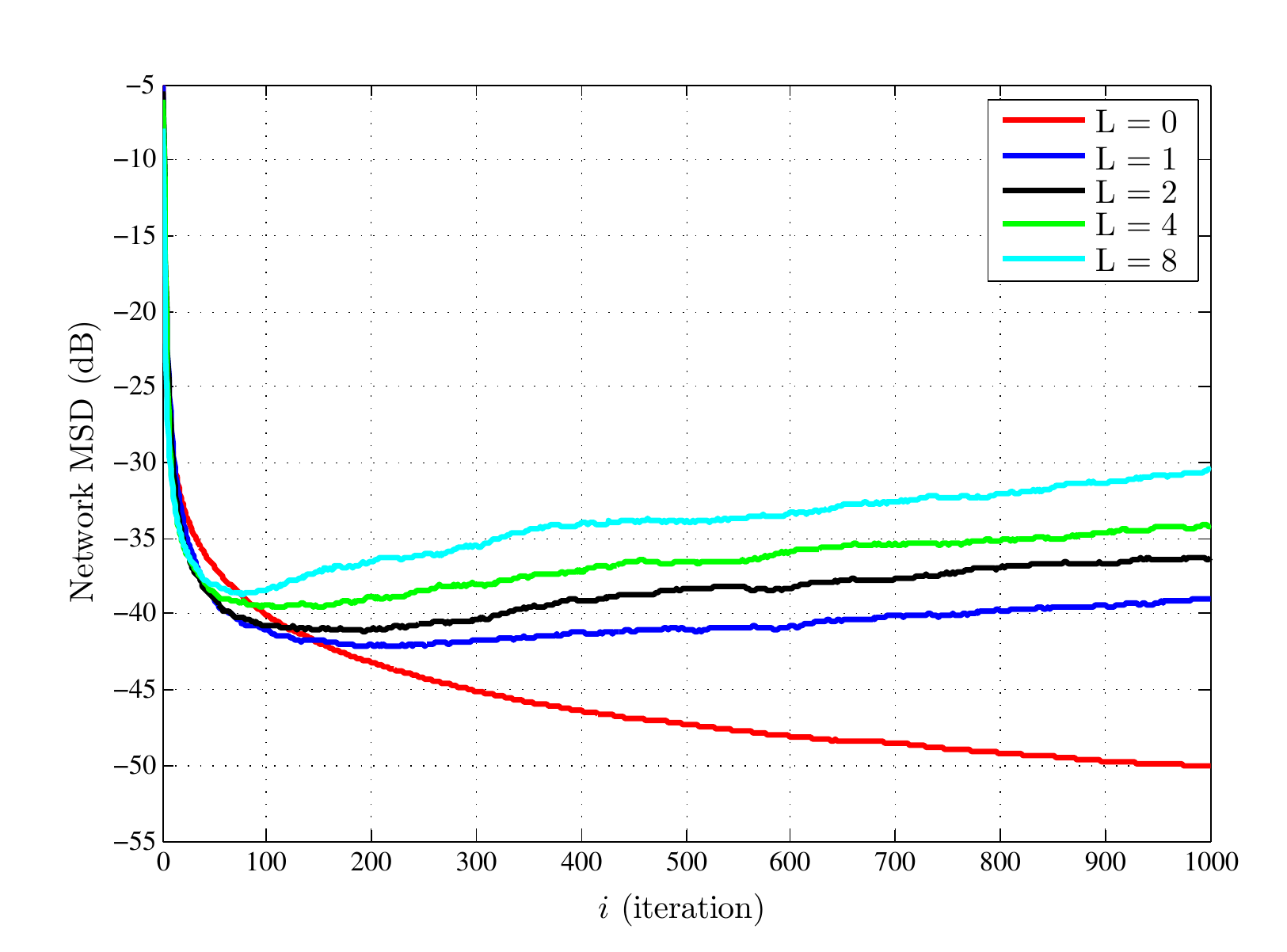}}   
  \caption{Simulated network MSD curves for partial diffusion RLS algorithms using sequential (top) and stochastic (bottom) with different number of entries communicated under noisy links $\lambda=1$.}
 \end{figure}
%%%%%%%%%%%%%%%%%%%%%%%%%%%%%%%%%%%%%%%%%%%%%%%
%%%%%%%%%%%%%%%%%%%%%%%%%%%%%%%%%%%%%%%%%%%%%%%
\begin{figure}  [t]% fig 10
  \centering
  \subfigure{\label{figtop:5}\includegraphics[width=8.5cm]{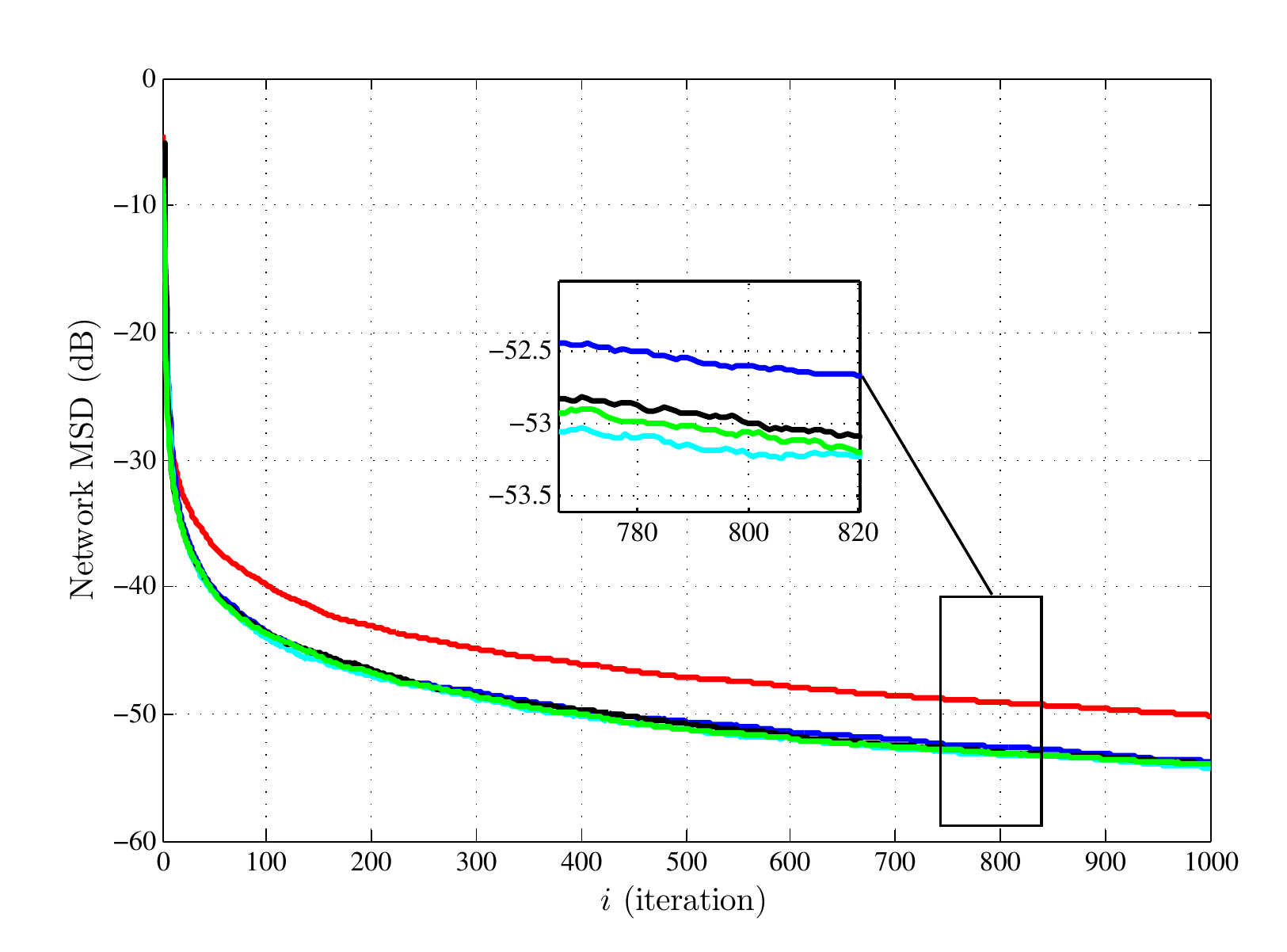}}  
  \subfigure{\label{figbottom:5}\includegraphics[width=8.5cm]{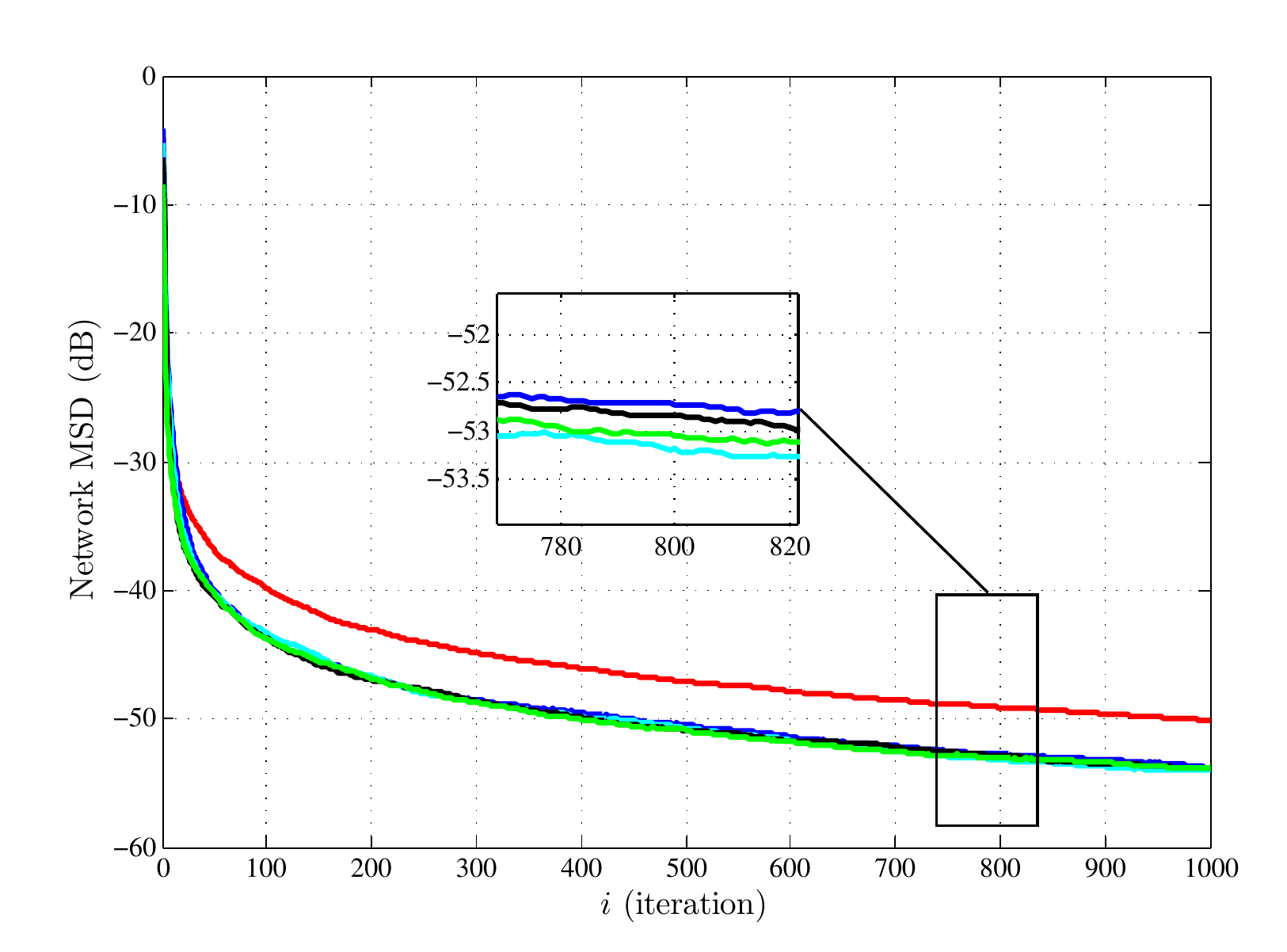}}   
  \caption{Simulated network MSD curves for partial diffusion RLS algorithms using sequential (top) and stochastic (bottom) with different number of entries communicated under ideal links when $\lambda=1$.}
 \end{figure}
%%%%%%%%%%%%%%%%%%%%%%%%%%%%%%%%%%%%%%%%%%%%%%%

\section{Conclusion and Future Work}
In this work, we investigated the performance of PDRLS algorithm when links between nodes were noisy. We derived an analytical expression for the network mean-square deviation, MSD, using weighted energy conservation relation. Our results revealed that the noisy links are the main factor in the performance degradation of PDRLS algorithm. They also illustrated that the stability conditions for PDRLS under noisy links are not sufficient to guarantee its convergence. In other words, considering non-ideal links condition added a new complexity to the estimation problem for which the PDRLS algorithm became unstable and did not converge for any value of forgetting factor, $\lambda$. It was also showed that the PDLMS algorithm with noisy links shows divergent behavior for both selection schemes (stochastic and sequential), while it does not for the noise-free links case. 
In the future, tighter and accurate bounds on the convergence rate of the mean and mean-square update equations of PDRLS algorithm with noisy links can be established. Necessary and sufficient conditions for convergence of the algorithm with noisy links need to be derived. These can be addressed in the future.

\end{document}